\documentclass[reprint,superscriptaddress,preprintnumbers,amsmath,amssymb,aps,prl]{revtex4-2}

\usepackage{graphicx}
\usepackage{dcolumn}
\usepackage{bm}
\usepackage[mathlines]{lineno}

\usepackage{xcolor}

\usepackage{scalerel}

\newcommand{\uli}{\underline}

\usepackage{hyperref}

\makeatletter
\AtBeginDocument{%
  \def\toclevel@subsection{0}%
  \def\toclevel@subsubsection{1}%
}
\makeatother

\usepackage{scalerel}
\usepackage{relsize} 
\usepackage{etoolbox} 
\usepackage{relsize}   
\let\oldint\int

\renewcommand{\int}{\mathop{\scaleobj{0.85}{\oldint}}\nolimits}

\usepackage{ragged2e}
\usepackage{caption}

\begin{document}

\preprint{MPP-2026-178}

\title{A Quantum-Mechanical Model for the M5-Brane}

\author{Manuel Artime}
\affiliation{Max-Planck-Institut f\"ur Physik (Werner-Heisenberg-Institut),\\ 
Boltzmannstra\ss e  8,  85748 Garching, Germany\vspace*{5pt}}

\author{Ralph Blumenhagen}
\email{blumenha@mpp.mpg.de} 
\affiliation{Max-Planck-Institut f\"ur Physik (Werner-Heisenberg-Institut),\\ 
Boltzmannstra\ss e  8,  85748 Garching, Germany\vspace*{5pt}}

\author{Thomas Raml}
\affiliation{Max-Planck-Institut f\"ur Physik (Werner-Heisenberg-Institut),\\ 
Boltzmannstra\ss e  8,  85748 Garching, Germany\vspace*{5pt}}
\affiliation{Arnold Sommerfeld Center for Theoretical Physics,\\
Ludwig-Maximilians-Universit\"at, 80333 M\"unchen, Germany}

\setlength{\abovedisplayskip}{6pt}
\setlength{\belowdisplayskip}{6pt}

\begin{abstract}
We present a maximally supersymmetric extension of the BFSS matrix model based on the 5-Lie algebra ${\mathcal A}_6$. The theory is Lagrangian, promotes the BFSS 2-bracket structure constants to dynamical objects governed by a Chern-Simons-like kinetic term, and displays features expected from a theory of M5-branes. We show that it can also be obtained from the single lightcone M5-brane action on a spatial five-sphere after a truncation to the time coordinate. The resulting theory is a consistent truncation and, analogously to the $U(2)$ BFSS model, supports an interpretation of the $\mathcal{A}_6$ theory as a quantum-mechanical model for the lowest-lying modes of the M5-brane.
\end{abstract}

\maketitle


\subsection*{Introduction}\label{sec:intro}

\vspace{-1em}
A unified description of quantum mechanics and gravity remains a central problem in fundamental physics. Accordingly, a microscopic formulation of quantum gravity must not only explain how spacetime emerges from quantum degrees of freedom, but also how extended objects that probe its nonperturbative structure are encoded in such a description. Matrix quantum mechanics provides a concrete setting in which these questions can be addressed while maintaining computational control.

String theory remains the most developed candidate for such a theory of quantum gravity. Already in its perturbative description it is striking that spacetime is not fundamental but appears through background fields acting as couplings in the two-dimensional quantum field theory on the worldsheet swept out by the string. Beyond perturbation theory, dualities relate seemingly distinct string theories, revealing the presence of higher-dimensional objects, called branes, and point to an eleven-dimensional theory, M-theory.  About this theory far less is known, but it is believed to arise in the strong coupling limit of ten-dimensional type IIA string theory, where instead of strings, particle-like states, namely the type IIA D0-branes, arise as the lightest species. 

Banks, Fischler, Shenker, and Susskind (BFSS) conjectured~\cite{Banks:1996vh}
(see~\cite{Bilal:1997fy,Bigatti:1997jy,Taylor:2001vb} for reviews) that in  infinite momentum frame, M-theory can then be exactly described by the quantum mechanics on the worldline of $N$ D0-branes, given by the super Yang--Mills $U(N)$ action
\begin{align}\label{eq:actionbfss}
    S_{\rm BFSS} &= \tfrac{1}{2} \int  \mathrm{d}t \bigl(  D_t X^I_a D_t X^I_a - \tfrac{1}{2} f_{abc} f_{ade} X^I_b X^J_c X^I_d X^J_e\nonumber\\
    &\qquad \qquad + i \Theta_a^T D_t \Theta_a + i f_{abc} \Theta_a^T \Gamma^I \Theta_bX^I_c \bigr)\,,
\end{align}
by taking  $N$ to infinity. The same quantum mechanics had been obtained earlier by de Wit--Hoppe--Nicolai~\cite{deWit:1988wri} in 1988, when introducing a regularization of the supersymmetric (single) M2-brane in lightcone gauge, where the membrane had the topology of a spatial two-sphere.

\pagebreak
At the same time, M5-branes remain elusive in this model, as, for example, no transverse M5 charge appears in the supersymmetry algebra. This reflects the long-standing difficulty in formulating the six-dimensional theory on multiple M5-branes. While for transverse M5-branes this obstacle was overcome by a deformation of the matrix model~\cite{Berenstein:2002jq,Maldacena:2002rb}, they do not appear as explicit classical solutions of the matrix model and thus are not on the same footing as the M2-branes. However, the fact that the matrix regularization of a single M2-brane already gives an interacting theory suggests that a quantum-mechanical model, rather than the full-fledged six-dimensional worldvolume theory, may be a more accessible route toward multiple M5-branes.

Recently, in~\cite{Artime:2026fcr} we proposed a one-dimensional maximally supersymmetric action that extends the BFSS model by including a 5-bracket, naturally incorporating the classical  M5-brane. In general, maximally supersymmetric theories are rare and highly constraining.
Using maximal supersymmetry as a guiding principle, the mathematical consistency of the model led us to introduce additional structure. Specifically, we promoted the structure constants of the BFSS 2-bracket to a dynamical field $H_{abc}$ featuring a Chern-Simons-like kinetic term. This suggested an interpretation as the analog of the self-dual three-form on the worldvolume theory of the M5-brane. Although in general additional constraints on $H_{abc}$ were needed, these turn out to be absent by choosing the $\mathcal{A}_6$ 5-Lie  algebra with $F_{abcdef} = \frac{1}{3!}\epsilon_{abcdef}$.

After introducing the main features of the $\mathcal{A}_6$ theory, we present a classical solution describing a round M5-brane and show that the $\mathcal{A}_6$ extension arises as a consistent truncation of the M5-brane model of Bandos-Townsend~\cite{Bandos:2008fr} wrapped on a spatial $S^5$.
We confirm the role of $H$ as the quantum-mechanical realization of the three-form field and show that the self-duality condition is automatically implemented.
Moreover, we also propose a large self-dual-flux scaling limit, in which the $\mathcal{A}_6$ model reduces to the BFSS model with gauge algebra $\mathfrak{su}(2)\oplus \mathfrak{su}(2)$. 

\subsection{The \texorpdfstring{$\boldsymbol{\mathcal{A}_6}$}{A6}-extended BFSS model}\label{sec:A6}

\vspace{-1em}
Following closely our recent work~\cite{Artime:2026fcr} to manifestly include the M5-brane in the BFSS matrix model, we add a term containing a nontrivial 5-bracket to the bosonic action, i.e. schematically
\begin{align}
    S &= \tfrac{1}{2} \int \mathrm{d}t \,\mathrm{Tr}\bigl[ D_t X^I D_t X^I +\tfrac{1}{2} [X^I,X^J]^2\nonumber\\
   &\qquad \qquad  + \tfrac{1}{5!} [X^I,X^J,X^K,X^L,X^M]^2\,\bigr]\,.
\end{align}
Here, the antisymmetric 5-bracket is defined as
\begin{equation}
    [X^I\!,X^J\!,X^K\!,X^L\!,X^M]_f\!=\!F_{abcdef} X^I_a X^J_b X^K_c X^L_d X^M_e\!\,,
\end{equation}
by use of, a priori undetermined, totally antisymmetric ``structure
constants" subject to the so-called fundamental identity
\begin{align}\label{eq:FI}
    F_{[\underline{abcde}}{}^k\, F_{k\underline{f}] lmnp} &=0\,,
\end{align}
where only the underlined indices are antisymmetrized. This fundamental identity can be solved by choosing  the 5-Lie algebra $\mathcal{A}_6$,  for which $F_{abcdef} =\tfrac{1}{3!}\epsilon_{abcdef}$ with $a=1,\dots,6$ such that 
\begin{align}\label{eq:square_F}
    F_{abcdef}F_{a'b'c'def}=\delta_{[\underline{a}}^{a'}\delta_{\underline{b}}^{b'}\delta_{\underline{c}]}^{c'}\,,
\end{align}
which was also a key property in the more general framework of~\cite{Artime:2026fcr}. Furthermore, the epsilon tensor satisfies the Schouten identity
\begin{equation}
    \delta_{n[\underline{m\vphantom{f}}}F_{\underline{abcdef}]}=0\,,
\end{equation}
which crucially enters many of the explicit calculations.
It is  convenient to define a two-index ``adjoint'' gauge field $\mathcal{A}_{ab}$ and resulting covariant derivatives 
\begin{equation}
\begin{aligned}
    D_t X^I_a &= \partial_t X^I_a - \mathcal{A}_{ab}X^I_b\,,\\
    D_t \Theta_a &= \partial_t \Theta_a - \mathcal{A}_{ab}\Theta_b\,,\\
    D_t H_{abc} &= \partial_t H_{abc} - 3\mathcal{A}_{[\underline{a} d} H_{d \underline{bc}]}\,,
\end{aligned}
\end{equation}
as well as gauge transformations $\Lambda_{ab}(t)$ acting as
\begin{align}
    \delta_{\Lambda} X^I_a &= \Lambda_{ak} X^I_k\,, &
    \delta_{\Lambda} \Theta_a &= \Lambda_{ak} \Theta_k\,,\\
    \delta_{\Lambda} H_{abc} &=3\Lambda_{[\underline{a} k}H_{k \underline{bc}]}\,, & \delta_{\Lambda} \mathcal{A}_{ab} &= \partial_t \Lambda_{ab} + 2\Lambda_{[\underline{a}k} \mathcal{A}_{k\underline{b}]}\,.\nonumber
\end{align}
Here, $H_{abc}=H_{abc}(t)$ is the dynamical analog of the structure constants defining the 2-bracket in the standard BFSS model, i.e.
\begin{equation}
    f_{abc} \to H_{abc}(t)\,.
\end{equation}
With these definitions, one can then show that the action
\begin{align}
  \label{eq:A6_action}
    S_{\mathcal{A}_6} &=\tfrac{1}{2} \int \mathrm{d}t \bigl(D_t X^I_a D_t X^I_a + i \Theta_a^T D_t \Theta_a\nonumber\\
    & +i  H_{abc} \Theta_a^T \Gamma^I \Theta_bX^I_c- \tfrac{1}{2} H_{abc} H_{ade} X^I_b X^J_c X^I_d X^J_e\nonumber\\
    &-\! \tfrac{i}{4!}  F_{abcdef} \, \Theta_a^T \Gamma^{IJKL}\Theta_b  \, X^I_c X^J_d X^K_e X^L_f\nonumber\\
    &-\! \tfrac{1}{5!} F_{abcdef} F_{aghuvw} X^I_b\! X^J_c\! X^K_d\!  X^L_e\! X^M_f\! X^I_{g}\! X^J_{h}\! X^K_{u}\! X^L_{v}\! X^M_{w}\nonumber\\
    &+H_{abc}\, F_{abcdef}\, D_t H_{def}\bigr)\,,
\end{align}
is invariant under the above gauge transformations, as well as dynamical supersymmetry transformations. 

\noindent
The associated supersymmetry transformations read
\begin{equation}
\begin{aligned}
    \delta_\epsilon X^I_a &= -i \epsilon^T \Gamma^I \Theta_a\,,\\
    \delta_\epsilon \Theta_a &= \bigl( D_t X^I_a \Gamma^I + \tfrac{1}{2} H_{abc} X^I_b X^J_c\, \Gamma^{IJ}\\
    &+ \tfrac{1}{5!} F_{abcdef} X^I_b X^J_c X^K_d X^L_e X^M_f \, \Gamma^{IJKLM} \bigr) \,\epsilon\,,\\
    \delta_\epsilon \mathcal{A}_{ab} &= -i \epsilon^T H_{abc}\Theta_c\\
    & + \tfrac{i}{3!} \epsilon^T  F_{abcdef}   X^I_d X^J_e X^K_f\Gamma^{IJK} \Theta_c\,,\\
    \delta_\epsilon H_{abc} &= -\tfrac{1}{2}i F_{abcdef} (\epsilon^T \Gamma^{JK} \Theta_d) X^J_e X^K_f\,,
\end{aligned}
\end{equation}
where $\epsilon$ is a constant 16-component Majorana spinor.\newline
\indent Using these transformations, one can then also check that the supersymmetry algebra closes on-shell, i.e. on gauge transformations and equations of motion
\begin{align}
    \relax [\delta_{\epsilon_2},\delta_{\epsilon_1}] X^I_a &= 2i \, (\epsilon_2^T \epsilon_1) \,D_t X^I_a + \delta_{\Lambda} X^I_a\,,\nonumber\\
    [\delta_{\epsilon_2},\delta_{\epsilon_1}] H_{abc} &=
    2i(\epsilon_2^T \epsilon_1) D_t H_{abc} +\delta_{\Lambda} H_{abc} + \mathrm{e.o.m.}\,,\nonumber\\
    [\delta_{\epsilon_2},\delta_{\epsilon_1}] {\Theta}_{a} &= 2i(\epsilon_2^T \epsilon_1) D_t \Theta_a + \delta_{\Lambda}\Theta_a + \mathrm{e.o.m.}\,,\nonumber\\
 [\delta_{\epsilon_2},\delta_{\epsilon_1}] \mathcal{A}_{ab} &= \delta_{\Lambda} \mathcal{A}_{ab} + \mathrm{e.o.m.}\,,
\end{align}
with common gauge parameter
\begin{equation}
\begin{aligned}
    \Lambda_{ab}&=2i(\epsilon_2^T\Gamma^I\epsilon_1)H_{abc}X^I_c\\
    &\qquad +2i(\epsilon_2^T\Gamma^{IJKL}\epsilon_1)\tfrac{1}{4!}F_{abcdef}X^I_cX^J_dX^K_eX^L_f\,.
\end{aligned}
\end{equation}
As in~\cite{Artime:2026fcr}, one can add a zero component to the index set so that after setting  $\mathcal{A}_{0a}=0$, $H_{0ab}=0$ and $F_{0abcde}=0$, the action also becomes invariant under 16 so-called kinematic supersymmetries, only shifting the zero component $\Theta_0$. Hence, as for the BFSS model, in total, we have 32 supersymmetries.

\vspace{-1em}
\subsection{M5-branes in the  \texorpdfstring{$\boldsymbol{\mathcal{A}_6}$}{A6}-extended  model}

\vspace{-1em}
In this section, we provide strong evidence that the $\mathcal{A}_6$ model provides an exact quantum-mechanical description of the lowest modes of a single M5-brane wrapping $S^5$. 

\vspace{-1em}
\subsubsection{A classical round  M5-brane solution}

\vspace{-1em}
Equipped with a Lagrangian theory, one can analyze the classical equations of motion, which read
\begin{align}\label{eom:A6}
    0&=D_t^2X^I_a+H_{abc}H_{bde}X^J_cX^J_d\,X^I_e-\tfrac{i}{2} H_{abc}\Theta_b^T\Gamma^I\Theta_c\nonumber\\
    &+\tfrac{i}{2\cdot 3!}F_{abcdef} \Theta^T_b\Gamma^{IJKL}\Theta_c X^J_dX^K_eX^L_f\nonumber\\
    &-\tfrac{1}{4!}F_{abcdef} F_{bghuvw} X^J_cX^K_dX^L_e X^M_fX^J_g X^K_h X^L_uX^M_vX^I_w\,,\nonumber\\
    0&=D_t\Theta_a+H_{abc}\Gamma^I\Theta_bX^I_c\\
    &-\tfrac{1}{4!}F_{abcdef}\Gamma^{IJKL}\Theta_bX^I_cX^J_dX^K_eX^L_f\,,\nonumber\\
    0&=D_tH_{abc}-\tfrac{i}{2}F_{abcdef}\Theta^T_d\Gamma^I\Theta_e X^I_f\nonumber\\
    &+\tfrac{1}{2}F_{abcdeh}H_{fgh}X^I_dX^J_e\,X^I_fX^J_g\,,\nonumber
\end{align}
together with the constraint 
\begin{equation}\label{eq:Gauss}
      0=X^I_{[a} D_tX^I_{b]}-\tfrac{i}{2}\Theta_{[a}^T\Theta_{b]}-\tfrac{3}{2}H_{efg}F_{efg[\underline{a}cd}H_{\underline{b}]cd}\,,
\end{equation}
which arises from $\mathcal{A}$ appearing in the action as a Lagrange multiplier and can be viewed as the generalization of the so-called Gauss constraint in the standard BFSS model.

We want to find a solution to these equations with vanishing fermions $\Theta_a=0$, and $X^I$ that describe a round $S^5$ with isotropic (and hence $SO(6)$-invariant) $H^2$
\begin{equation}
    X^I_a X^I_b \sim \delta_{ab}\,, \qquad  H_{acd}H_{b}{}^{cd} \sim \delta_{ab}\,.
\end{equation}
This can be done in a systematic manner utilizing the $SU(3)$-structure of the embedding space $V=\mathbb{R}^6\cong\mathbb{C}^3$. Here, we just report on one such solution which reads
\begin{align}
    X_a^I&=r_0 \,\delta_a^I\,,\qquad\
    \mathcal{A}= -\tfrac{5}{6}r_0^4\,\omega\,,\\
    H(t)&= \pm\tfrac{\sqrt{5}}{6}r_0^3 \left(\cos(3r_0^4 t){\rm Re}(\Omega) + \sin(3r_0^4 t){\rm Im}(\Omega) \right)\,,\nonumber
\end{align}
where we used the canonical K\"ahler form
${\omega=\sum_{i=1}^3 {\rm d}y_{2i-1}\wedge {\rm d}y_{2i}}$ and the holomorphic
three-form ${\Omega=\wedge_{i=1}^3 ({\rm d}y_{2i-1}+ i {\rm d}y_{2i})}$. Note that while
the $X^I$ define a sphere of constant radius $r_0$, the threading
$H$-flux is not constant and the solution requires a nontrivial gauge field $\mathcal{A}$. This solution is not supersymmetric, but can be interpreted as an M5-brane wrapping a round static $S^5$ which is stabilized by the (rotating) $H$-flux and the nontrivial gauge field. This provides additional motivation to consider the worldvolume theory of an M5-brane wrapped on an $S^5$.

\subsubsection{The \texorpdfstring{$\mathcal{A}_6$}{A6} model from the lightcone M5 theory}

\vspace{-1em}
The action of the $U(N)$ BFSS matrix model was originally derived via a truncated dimensional reduction of the single three-dimensional supermembrane theory on $S^2$ in lightcone gauge. If one restricts to the linear harmonics, this leads to the BFSS model with gauge algebra $\mathfrak{su}(2)$. Naturally, one might ask if a similar relation holds for the $\mathcal{A}_6$ extension. In this section, we report that the latter arises from the single M5-brane theory in the lightcone gauge of Bandos-Townsend~\cite{Bandos:2008fr} by wrapping the M5-brane on a five-sphere and performing a truncated dimensional reduction to a single time coordinate, retaining only the modes linear (and constant) in the embedding coordinates. This gives a consistent truncation, such that all solutions of the $\mathcal{A}_6$ model uplift to solutions of the M5 worldvolume theory.

Moreover, for this model to be a generalization of the BFSS matrix model, one needs to clarify how the latter is embedded. In the appendix, we present a large $H$-flux scaling limit that
indeed recovers the BFSS model.

In~\cite{Bandos:2008fr}, the Hamiltonian formalism of~\cite{Bergshoeff:1998vx} was employed in which the worldvolume of the M5-brane splits into $(t, \sigma^i)$, where the $\sigma^i$ parameterize the five-dimensional spatial part. The coordinates $X^I(t, \sigma)$ describe the embedding of the M5-brane into the nine-dimensional flat background, giving an induced metric on the spatial part of the worldvolume $(g_5)_{ij} = \partial_i X^I \partial_j X^I$. 
Moreover, the theory contains 16-component real $SO(9)$ spinors $\theta_+(t, \sigma)$, and a three-form flux $H = \mathrm{d}B - C$. The latter is constructed from the two-form $B$ living on the M5-brane worldvolume and the three-form bulk potential $C$.\pagebreak

Our starting point for this six-dimensional theory,
after going to lightcone gauge, is the following Lagrangian~\cite{Bandos:2008fr}
\begin{align}\label{eq:M5_LC_second_order_sign}
    &S_{\rm BT}= \int {\rm d}t\, {\rm d}^5\sigma \Big( \tfrac{\mu_0}{4}D_t X^I D_t X^I +\tfrac{i \mu_0}{2}\,\theta_+^TD_t\theta_+ \nonumber\\
    &+\tfrac{i}{2} \epsilon^{ijklm} \partial_iX^J(\partial_j \theta^T_+ \Gamma^J \theta_+) \partial_k B_{lm} -\tfrac{|g_5|}{3! \mu_0}  H_{ijk} H^{ijk}\nonumber\\
    &+\tfrac{i}{24} \epsilon^{ijklm} \partial_iX^I  \partial_j X^J \partial_k X^K \partial_l X^L (\partial_m \theta^T_+
    \Gamma^{IJKL} \theta_+)\nonumber\\
    &-\tfrac{|g_5|}{\mu_0} -\tfrac{1}{4!}\epsilon^{ijklm} D_t B_{ij} H_{klm}\Big)\,.
\end{align}
Here, the fermionic signs are matched to our conventions, and the M5-brane tension is set to $T_{\rm M5}=1$. The quantity $\mu_0$ is the volume element of a fiducial metric $\mu = \mu_0\,\mathrm{d}^5 \sigma$. Note that in lightcone gauge and due to their Hamiltonian origin, only the spatial (magnetic) components of $B$ and $H$ appear in the action, while $B_{0i}$ only gives rise to surface terms~\cite{Bandos:2008fr}. The covariant derivatives read
\begin{align}\label{eq:defcovderivBT}
    D_t X^I &= \partial_t X^I + s^i \partial_i X^I\,, \quad D_t \theta_+=\partial_t \theta_+ + s^i \partial_i
    \theta_+\,,\nonumber\\
    D_t B_{ij} &= \partial_t B_{ij} + s^k H_{kij}\,,
\end{align}
with the vector field $s^i$ that is subject to the constraint 
\begin{equation}\label{eq:shift_constraint}
    \partial_i (\mu_0s^i)=0\,.
  \end{equation}
In fact, before arriving at the action~\eqref{eq:M5_LC_second_order_sign} a
second constraint appeared in~\cite{Bandos:2008fr}, namely 
\begin{align}\label{eq:secondconstraint}
  &\partial_{[i} K_{j]}=0\,,\\
K_{i} \!=\! \bigl(\tfrac{1}{2}\partial_i X^I \!D_t X^I \!- &\tfrac{1}{4! \mu_0} \epsilon^{jklmn}H_{ijk}H_{lmn} \!-\! \tfrac{i}{2}\partial_i \theta^T_+ \theta_+\bigr)\,,\nonumber
\end{align}
such that together these constraints implemented the spatial diffeomorphism invariance, which in the lightcone Hamiltonian description is broken to 5-volume-preserving diffeomorphisms. The theory therefore has an exotic gauge invariance with respect to $\mathrm{SDiff}(\Sigma_5)$. For further details and conventions, we refer to~\cite{Bandos:2008fr}.

The canonical Nambu-Poisson 5-bracket for functions $f_i \in C^\infty(\Sigma_5)$ is defined via the Jacobian determinant
\begin{equation}\label{eq:Nambu}
\{f_1, f_2, f_3, f_4, f_5\} = \frac{\epsilon_{ijklm}}{\mu_0}
\partial_i f_1 \partial_j f_2 \partial_kf_3
\partial_l f_4 \partial_m f_5 \,,
\end{equation}
and turns the vector space $C^\infty(\Sigma_5)$ into a Nambu-Poisson algebra. Importantly, the above bracket then satisfies the fundamental identity~\eqref{eq:FI} and hence provides an example of a infinite-dimensional 5-Lie algebra. Its (Nambu-)Hamiltonian vector fields are divergence free and therefore gauge theories associated with this bracket possess the residual gauge symmetry $\mathrm{SDiff}(\Sigma_5)$.
Now, we choose the spatial part of the M5-brane to be the round $S^5 \subset \mathbb{R}^6$, for which  we have the standard spherical embedding coordinates $y_a$, $a=1,\dots 6$ satisfying  $\sum_a (y_a)^2=1$. Moreover, the volume form satisfies
\begin{equation}\label{eq:conventionsdimred}
    V_5 = \int_{S^5} \mu\,, \qquad I_{ab}= \int_{S^5} \mu\, y_a y_b = \tfrac{V_5}{6}\delta_{ab}\,,
  \end{equation}
so that for the embedding coordinates one gets
\begin{equation}\label{eq:5bracket_y}
\{y_a, y_b, y_c, y_d, y_e\} = \epsilon_{fabcde} y_f \,.
\end{equation}
Denoting $\mathcal{Y}=\mathrm{span}\{y_a\}$, we see that $(\mathcal{Y}, \{\cdot, \cdot, \cdot,\cdot, \cdot \}_5)$ gives rise to the 5-Lie sub-algebra $\mathcal{A}_6 \subset C^\infty(S^5)$.

We can now perform the dimensional reduction of the action~\eqref{eq:M5_LC_second_order_sign} to the time-direction while  truncating the internal dependence to the modes
in  $\mathcal{Y}$. This means that we expand the fields in terms of these modes as
\begin{align}\label{eq:modeexpand}
    X^I(t,\sigma) &= c_X X^I_a(t)\,y_a(\sigma)\,, \quad
    \theta_+(t,\sigma) = c_\Theta \Theta_a(t)\,y_a(\sigma) \,,\nonumber\\
    B_{ij}(t,\sigma)&=\tfrac{c_H}{3}  \epsilon_{abcdef} H_{abc}(t) y_d \partial_i y_e \partial_j y_f\,,  
\end{align}
and $H_{ijk}=3\partial_{[k} B_{ij]}$ with normalization parameters $c_X$, $c_\Theta$ and  $c_H$. Next, we need to define the vector field appearing in the covariant derivative. A convenient choice are the $SO(6)$ Killing vectors
\begin{equation}\label{eq:Killing}
    \mathcal{K}^i_{ab}(\sigma) = \gamma^{ij}(y_a \partial_j y_b - y_b \partial_j y_a)\,,\quad \gamma_{ij}=\partial_i y_a \partial_j y_a\,,
\end{equation}
which are themselves divergence-free and define $s$ via
\begin{equation}\label{eq:vector_s}
\begin{aligned}
    s^i(t,\sigma)= -\tfrac{1}{2}\mathcal{A}_{ab}(t)\,\mathcal{K}^i_{ab}(\sigma)\,,
\end{aligned}
\end{equation}
with antisymmetric coefficient $\mathcal{A}_{ab}(t)$, which guarantees that $s^i\partial_i y_a={\cal A}_{ab} y_b$. This implies that the
covariant derivative becomes $D_t X^I=  c_X D_t X^I_a y_a$ and lies again in the truncated space $\mathcal{Y}$.
Moreover, this choice automatically satisfies the constraint~\eqref{eq:shift_constraint}.

Now we are ready to perform the dimensional reduction. This is a tedious, but in principle straightforward computation. We present some core steps and formulas in the appendix. At the end of the day, we obtain
\begin{equation}
    S^{\rm trunc}_{\rm BT}=\kappa\, S_{{\mathcal A}_6}\,,
\end{equation}
where we have to fix the normalization constants to
\begin{equation}\label{eq:normal_werder}
       c_X=\frac{1}{12^{1/4}}\,,\quad c_\Theta=\frac{c_X}{\sqrt{2}}\,,\quad   c_H=c_X^3
\end{equation}
and $\kappa=V_5/(12)^{3/2}$. Here, the correspondence is really term by term in the order the two actions were presented in~\eqref{eq:A6_action} and~\eqref{eq:M5_LC_second_order_sign}.
Notice that the constraint~\eqref{eq:secondconstraint} for the lightcone M5 action is nothing but the generalized Gauss constraint~\eqref{eq:Gauss} in the ${\mathcal A}_6$ model.

\begin{figure}[t]
\includegraphics[width=1\linewidth]{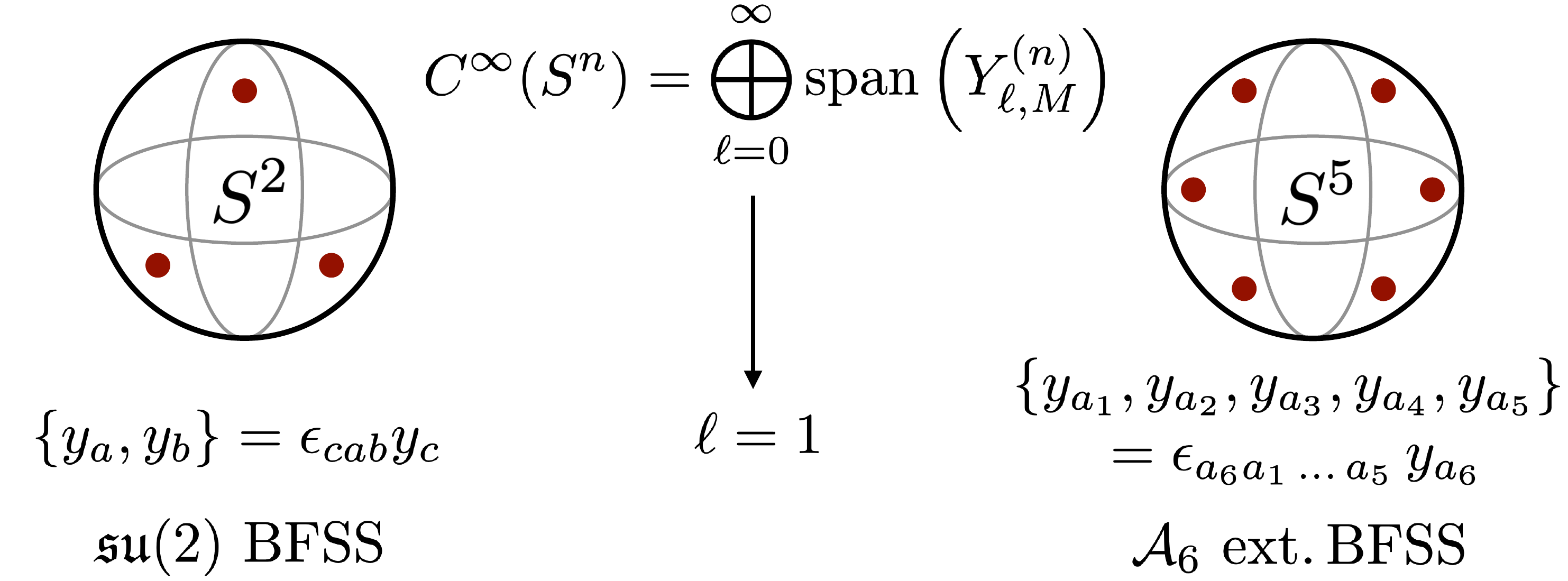}
\caption{\label{fig:truncation} 
Similar to the BFSS model, the $\mathcal{A}_6$ model arises from the truncated reduction of the M5-brane action on $S^5$.}
\end{figure}

Hence, this correspondence provides a precise match, rendering the picture  completely analogous to the BFSS $SU(2)$ matrix model. The latter relates to the single supermembrane theory on $S^2$ dimensionally reduced to the time direction, retaining only the modes linear in the embedding coordinates, see also Figure~\ref{fig:truncation}. Analogously, one can add the single constant mode that extends the BFSS matrix model to $U(2)$ and, in our case, adds a zero component to the index  set, as detailed above and in~\cite{Artime:2026fcr}.

\vspace{-1em}
\subsubsection{Self-duality of the three-form field}

\vspace{-1em}
A central requirement of any M5-brane description is to incorporate the self-duality of the three-form $H$. In fact, this  was a common objection to any higher-bracket implementation of the M5-brane
within matrix models.
The present reduction reveals that this property is in fact already implemented  in the $\mathcal{A}_6$ model.

In order to see this, it is first instructive to calculate the equations of motion for $B_{ij}$, which can be written as
\begin{align}\label{eq:eom_B}
    0 = \tfrac{1}{4}\epsilon^{ijklm} \partial_k\bigl( D_t B_{lm} +  \tfrac{2\sqrt{|g_5|}}{\mu_0} (\star_5 \widehat{H})_{lm}) -  C_{0lm} \bigr)\,,
\end{align}
with $\widehat{H}$ the spatial part $H_{ijk}$ of the full six-dimensional three-form $H$ and the Hodge star operator $\star_5$ is defined with respect to $g_5$. Truncating to the linear modes gives
\begin{equation}
    0 = \tfrac{c_H}{3!}\epsilon_{abcdef}\partial_i y_d \partial_j y_e \partial_k y_f ( \mathrm{e.o.m.}_{H})_{abc}\,,
\end{equation}
with $\mathrm{e.o.m.}_{H}$ the equation of motion of $H$ in the $\mathcal{A}_6$ theory given in~\eqref{eom:A6}. Therefore, any solution of $H$ in the $\mathcal{A}_6$ theory lifts to a solution of $B_{ij}$ under our truncation. 

At the same time, equation~\eqref{eq:eom_B} can be expressed as
\begin{equation}\label{eq:self_duality}
    \mathcal{E}_2 = -\tfrac{2\sqrt{|g_5|}}{\mu_0} \ast_5 \widehat{H}\,,
\end{equation}
with $(\mathcal{E}_2)_{ij}= D_t B_{ij} - C_{0ij} - 2 \partial_{[i} B_{0j]}$ the covariant electric part of $\mathcal{H} = \mathrm{d}t  \wedge  \mathcal{E}_2 + \widehat{H}$. This is the nonlinear self-duality equation relating the (covariantized) electric and magnetic parts of $\mathcal{H}$, cf. also~\cite{Bergshoeff:1998vx}. Since $B_{0i}$ was unconstrained in the reduction and $H^2(S^5)=0$, for any solution $H_{abc}$, we can always find a $B_{0i}$ such that~\eqref{eq:eom_B} and~\eqref{eq:self_duality} are satisfied. Linearizing around a constant bosonic background with vanishing background flux and fixing the gauge to $s^i=0$, this leads to the familiar
\begin{align}
    \bar \ast_{1,5} h = h\,,
\end{align}
with $\bar \ast_{1,5}$ corresponding to the constant background metric and $h$ the linearized part of $H$.

\vspace{-1em}
\subsubsection{Consistent truncation}

\vspace{-1em}
For the theory to be a consistent truncation, one must check that the discarded higher harmonics are not sourced, i.e. that the equations of motion of \eqref{eq:M5_LC_second_order_sign} evaluated on the ansatz~\eqref{eq:modeexpand}--\eqref{eq:vector_s} have no component along them. This is indeed the case, as every term reduces to a Nambu-Poisson 5-bracket which, under the ansatz, closes by \eqref{eq:5bracket_y}. Importantly, the constraint \eqref{eq:secondconstraint} must also be consistent with the ansatz. One can easily check that
\begin{equation}
    K_i|_{\mathrm{trunc.}}=\tfrac{c_X^2}{2}C_{ab}y_b\partial_i y_a + \partial_i \Phi\,,
\end{equation}
with $C_{ab}\!=\!C_{[ab]}$ independent of higher harmonics and given by the generalized Gauss constraint \eqref{eq:Gauss}, thus, imposing $\partial_{[i} K_{j]}|_\mathrm{trunc.}=\partial_{[i} \partial_{j]} \Phi=0$.

\vspace{-1em}
\subsection{Conclusions}

\vspace{-1em}
By promoting the structure constants of the original BFSS matrix model
to a dynamical field, we unveiled a novel maximally supersymmetric
extension based on the $\mathcal{A}_6$ 5-Lie algebra. After presenting
a round $S^5$ solution, we showed that the very same action can be
obtained via a truncated dimensional reduction of the Bandos-Townsend
lightcone M5-brane action. The truncation is in fact consistent and
hence  $\mathcal{A}_6$ can be seen as an exact quantum-mechanical
model for the lowest-lying modes of an M5-brane wrapped on $S^5$ in
lightcone gauge. Furthermore, the theory  already implements the notorious self-duality of the three-form $H$.

Comparing this with the work of de Wit--Hoppe--Nicolai, it is clear that at
present, there is no obvious parameter $N$ that allows for a full
regularization procedure. This raises the question of whether this model
is an isolated case or,  similar to BLG~\cite{Bagger:2007jr,Gustavsson:2007vu} and ABJM~\cite{Aharony:2008ug}, part of a broader class
of models yet to be explored. In particular, as suggested
in~\cite{Artime:2026fcr}, a similar extension might also exist for the
M6 and M9-branes. This signals that this is only a first step and
that further extending the theory remains an open task.

\vspace{-1em}
\subsection{Acknowledgments}

\vspace{-1em}
We thank Quang-Khanh Pham and Matteo Zatti for insightful comments on the draft. M.A. would like to thank the organizers and participants of the workshop ``Pro(v/b)ing the Swampland'' at IFT Madrid for valuable discussions and kind hospitality during the final stages of this work. The work of R.B. is supported  by the Deutsche Forschungsgemeinschaft (DFG, German Research Foundation) under Germany’s Excellence Strategy – EXC-2094 – 390783311.

\bibliography{apssamp}

\appendix

\vspace{-1em}
\subsection{The BFSS limit}

\vspace{-1em}
For the $\mathcal{A}_6$ theory to be an extension of the BFSS model, one has to clarify how the latter is contained in it. Similarly to how a D0-brane can be constructed from the D2-brane with gauge flux by taking an appropriate large constant flux limit, we consider a pure classical $H$-flux background $H_{abc}=f_{abc}$. Then with all other fields except $\mathcal{A}$ vanishing, the Gauss constraint reduces to
\begin{equation}
    f_{efg}F_{efg[\underline{a}cd}f_{\underline{b}]cd} = -\tfrac{1}{4} \epsilon_{abcdef} f_{k[\underline{cd}} f_{k\underline{ef}]} =0\,,
\end{equation}
for all pairs of indices $a,b$. This vanishes if and only if the Jacobi-identity for $f_{abc}$ is satisfied. The equation of motion for $H_{abc}$  also implies $D_t f_{abc}=0$, which for a constant flux means ${\mathcal A}_{[\underline{a}d} f_{d\underline{bc}]}=0$. Therefore  $\mathcal{A}$ acts as a derivation and it follows that for constant $f_{abc}$, we can write the usual adjoint action ${\cal A}_{ab}=f_{abc} A_c$.

Then, taking the scaling limit $f_{abc}\sim \lambda f_{abc}\to \infty$ we can consider the theory for fluctuations (denoted in lower case) of the fields around this classical vacuum
\begin{equation}
\begin{aligned}
    X^I_a &= \lambda^{-1}x_a^I\, ,\quad \Theta_a = \lambda^{-1}\theta_a\,, \quad
      \mathcal{A}_{ab} =  \lambda^{0} a_{ab}\,,\\
    H_{abc} &= \lambda f_{abc} +  \lambda^{-3}  h_{abc}\,,
  \end{aligned}
\end{equation}
which is consistent with the supersymmetry variations.

Noticing that now the leading order CS-term is a total derivative, the dominant terms in the action~\eqref{eq:A6_action}, scaling as $\lambda^{-2}$, are precisely the terms of the BFSS action~\eqref{eq:actionbfss} with $f_{abc}$ the structure constants of a Lie algebra. However, in the present case, the only compact
semisimple choice is $\mathfrak{su}(2)\oplus \mathfrak{su}(2)$, which rather resembles the gauge algebra of the BLG model for multiple M2-branes than any $\mathfrak{su}(N)$ gauge algebra of the BFSS model. Including the constant mode adds a $\mathfrak{u}(1)$ factor.

\vspace{-1em}
\subsection{Details of the dimensional reduction}

\vspace{-1em}
In this appendix, we provide additional details on the main steps of the truncated dimensional reduction.

\vspace{-1em}
\subsubsection{The bosonic sector}

\vspace{-1em}
Let us start with the kinetic term for the $X^I$. Using the expansion~\eqref{eq:modeexpand} one can straightforwardly evaluate
\begin{align}
  \,\int_{\!\!\mathbb{R} \times S^5} &\mathrm{d}t\, \mathrm{d}^5\sigma\,
  \tfrac{\mu_0}{4}|D_t X^I|^2\nonumber\\
     &= \tfrac{c_X^2}{4} \int_{\!\!\mathbb{R} \times S^5} \mathrm{d}t\, \mathrm{d}^5\sigma\, \mu_0(D_t X^I_a)(D_t X^I_b)\,y^ay^b\nonumber\\
    &= \tfrac{c_X^2 V_5}{24}\int_{\!\!\mathbb{R}} \mathrm{d}t\, |D_t X^I_a|^2 =
   \tfrac{\kappa}{2} \int_{\!\!\mathbb{R}} \mathrm{d}t\, |D_t X^I_a|^2\,,
\end{align}
where in the third line we have used~\eqref{eq:conventionsdimred} and also implemented the normalizations~\eqref{eq:normal_werder}.

Next, we consider the term $|g_5|/\mu_0$ in $S_{\rm BT}$. First, one can express the determinant of the metric as
\begin{align}
    |g_5| &= \tfrac{1}{5!}\epsilon^{i_1i_2i_3i_4i_5}\epsilon^{j_1j_2j_3j_4j_5} g_{i_1 j_1}\dots g_{i_5j_5}\nonumber\\
    &=\tfrac{c_X^{10}}{5!} \epsilon^{i_1i_2i_3i_4i_5}\epsilon^{j_1j_2j_3j_4j_5} \partial_{i_1}y_a \!\!\dots \partial_{i_5} y_e \partial_{j_1}y_{a'} \!\!\dots \partial_{j_5}y_{e'} \!\times\nonumber\\
    &\quad\, X^I_a X^J_b X^K_c X^L_d X^M_e X^I_{a'} X^J_{b'} X^K_{c'} X^L_{d'} X^M_{e'}\,,
\end{align}
where we used the induced metric $(g_5)_{ij} = \partial_i X^I \partial_j X^I$.
Now realizing that these terms can be expressed via the 5-bracket relation~\eqref{eq:Nambu} as
\begin{align}
     \epsilon^{i_1i_2i_3i_4i_5}\partial_{i_1}y_a
     \partial_{i_2}y_b \partial_{i_3}y_c \partial_{i_4}y_d
     \partial_{i_5} y_e = \mu_0\epsilon_{fabcde} y_f\,,
\end{align}
it becomes straightforward to evaluate the integral as
\begin{align}
    \int_{\!\!\mathbb{R} \times S^5}\! &\mathrm{d}t\, \mathrm{d}^5\sigma\,\Bigl(-\tfrac{|g_5|}{\mu_0} \Bigr) \!=\! -\tfrac{\kappa}{2\cdot 5!} \! \int_{\!\!\mathbb{R}} \! \mathrm{d}t\, \tfrac{1}{3!} \epsilon_{abcdef}\,\tfrac{1}{3!}\epsilon_{a'b'c'd'e'f}\,\times\nonumber\\
    &\quad\,\,  X^I_a X^J_b X^K_c X^L_d X^M_e X^I_{a'} X^J_{b'} X^K_{c'} X^L_{d'} X^M_{e'}\,,
\end{align}
where again we used the normalizations~\eqref{eq:normal_werder}.
This indeed matches the corresponding term in $S_{\mathcal{A}_6}$.

In a similar vein one can evaluate the second bosonic potential term
\begin{align}
   & \int_{\!\!\mathbb{R} \times S^5} \mathrm{d}t\, \mathrm{d}^5\sigma\,  \left(-\tfrac{|g_5|}{\mu_0} \tfrac1{3!} H_{ijk} H^{ijk} \right) \nonumber\\
    &=-\tfrac{1}{3!} \int_{\!\!\mathbb{R} \times S^5}\mathrm{d}t\, \mu\, H_{ijk} H_{lmn}  \tfrac{|g|}{\mu_0^2} g^{il}g^{jm}g^{kn}\\
    &=-\tfrac{1}{2 (3!)^2} \int_{\!\!\mathbb{R} \times S^5}\mathrm{d}t\, \tfrac{\mu}{\mu_0^2}\, H_{ijk} H_{lmn}   \epsilon^{ijkop}\epsilon^{lmnqr}g_{oq}g_{pr}\,,\nonumber
\end{align}
where we have used the relation
\begin{equation}
    3! |g|g^{i[\uli{l}}g^{j\uli{m}}g^{k\uli{n}]} = \tfrac{1}{2}\epsilon^{ijkop}\epsilon^{lmnqr}g_{oq}g_{pr}\,.
\end{equation}
Inserting the expansion~\eqref{eq:modeexpand} for $H_{ijk}$ and noticing again
the appearance of the Nambu 5-bracket,  similar to the
case before one can   evaluate the integral as
\begin{align}
    \int_{\!\!\mathbb{R} \times S^5} \mathrm{d}t\, &\mathrm{d}^5\sigma\,\left(-\tfrac{|g_5|}{\mu_0}\tfrac{1}{3!} H_{ijk} H^{ijk}\right)\nonumber\\
    &=-\tfrac{\kappa}{4}\int_{\!\!\mathbb{R}}\mathrm{d}t\,H_{bca} H_{dea} X^I_b X^J_{c} X^I_{d} X^J_{e}\,.
\end{align}

From the bosonic terms it only remains to discuss the CS kinetic term, which requires some more work. It involves a term $D_t B_{ij}$, which is defined in~\eqref{eq:defcovderivBT}, and after some algebra can be expressed as
\begin{equation}
    D_t B_{ij} = (\partial_t + \mathcal{L}_s)B_{ij} - 2 \partial_{[i}Q_{j]}\,,
\end{equation}
where $\mathcal{L}_s$ denotes the usual Lie derivative along a vector
field $s$ and we defined $Q_{j}=s^kB_{kj}$. Noting also that
$\mathcal{L}_s(\partial_i y_a)= \mathcal{A}_{ab}\partial_iy_b$, after
some algebra one arrives at
\begin{equation}\label{eq:Dt_A_ij}
    D_t B_{ij} = \tfrac{c_H}{3}\epsilon_{abcdef}D_t H_{abc}y_d \partial_i y_e \partial_j y_f - 2 \partial_{[i}Q_{j]}\,.
\end{equation}
Now we are prepared to perform the dimensional reduction and evaluate the
CS-term in the Bandos-Townsend action. The first contribution from
\eqref{eq:Dt_A_ij}
proceeds analogously to the previous cases and leads to
\begin{align}
    \int_{\!\!\mathbb{R} \times S^5} &\mathrm{d}t\, \mathrm{d}^5\sigma\,\left(-\tfrac{1}{4!}\epsilon^{ijklm} D_t B_{ij} \,H_{klm}\right)\nonumber\\
   &= \tfrac{\kappa}{2} \int_{\!\!\mathbb{R}} \mathrm{d}t \,
     H_{abc}\,\tfrac{1}{(3!)}\epsilon_{abcdef}\, D_t H_{def}\,.
\end{align}
This is precisely the result expected from $\mathcal{A}_6$. It remains to consider the second contribution in~\eqref{eq:Dt_A_ij}. Let us focus on the appearing spatial integral, which we can write as
\begin{align}
    \int_{\!\! S^5} \mathrm{d}^5 \sigma \, & \epsilon^{ijklm} \partial_{[i}Q_{j]} H_{klm}\nonumber\\
    &= -\int_{\!\! S^5} \mathrm{d}^5\sigma \, \epsilon^{ijklm} Q_{j} \partial_{[i} H_{klm]}=0\,,
\end{align}
where we used that $\partial S^5=\emptyset$ and the Bianchi-identity following directly from the definition $H_{klm}=3\partial_{[k} B_{lm]}$. 

\vspace{-1em}
\subsubsection{The fermionic sector}

\vspace{-1em}
The kinetic term for $\theta_+$ is completely analogous to the kinetic term for the boson $X^I$. The kinetic term directly gives the desired
\begin{equation}
    -\!\int_{\!\!\mathbb{R} \times S^5} \!\!\mathrm{d}t\, \mathrm{d}^5\sigma\, \tfrac{i}2\mu_0\,(D_t\theta_+)^T\theta_+ = \tfrac{i \kappa}{2} \! \int_{\!\!\mathbb{R}} \!\! \mathrm{d}t\,\Theta_a^T D_t \Theta_a\,,
\end{equation}
while the fermionic Yukawa-like term yields
\begin{align}
    \frac{i}{2}  &\int_{\!\!\mathbb{R} \times S^5} \mathrm{d}t\, \mathrm{d}^5\sigma\, \epsilon^{ijklm} \partial_iX^J(\partial_j \theta^T_+ \Gamma^J \theta_+) \partial_k B_{lm}\nonumber\\
    &=\tfrac{i c_Xc_\Theta^2c_H}{6}\int_{\!\!\mathbb{R} \times S^5} \mathrm{d}t\, \mathrm{d}^5\sigma\, X_h^I \Theta^T_u\Gamma^I \Theta_v \epsilon_{abcdef}H_{abc} y_v\times \nonumber\\
    &\qquad\qquad\qquad\quad \big(\epsilon^{ijklm} \partial_i y_h \partial_j y_u \partial_k y_d \partial_l y_e \partial_m y_f\big) \,.
\end{align}
We see again the appearance for the Nambu 5-bracket so that  we can proceed as usual and evaluate the integral
\begin{align}
   \tfrac{i}{2}  &\int_{\!\!\mathbb{R} \times S^5} \mathrm{d}t\, \mathrm{d}^5\sigma\, \epsilon^{ijklm} \partial_iX^J(\partial_j \theta^T_+ \Gamma^J \theta_+) \partial_k B_{lm}\nonumber\\
 &=\tfrac{i\kappa}{2} \int_{\!\!\mathbb{R}} \mathrm{d}t\, H_{abc}  \Theta^T_a\Gamma^I \Theta_b X_c^I\,,
\end{align}
in agreement with the BFSS Yukawa term in $S_{\mathcal{A}_6}$.

\noindent For the term involving the 5-bracket, we finally obtain
\begin{align}
    &\!\!\!\tfrac{i}{24}\int_{\!\!\mathbb{R} \times S^5}\hspace{-12pt} \mathrm{d}t\, \mathrm{d}^5\sigma\epsilon^{ijklm} \partial_iX^I\! \partial_j X^J \!\partial_k X^K \!\partial_l X^L\!(\partial_m \theta^T_+ \Gamma^{IJKL} \theta_+\!)\nonumber\\
    &=-\tfrac{i\kappa}{2\cdot 4!}\int_{\!\!\mathbb{R}}\!\! \mathrm{d}t \, \tfrac{1}{3!}\epsilon^{abcdef}\Theta_a^T \Gamma^{IJKL}\Theta_b X_c^I X_d^J X_e^K X_f^L\,.\!
\end{align}

\end{document}